\documentclass[a4paper]{jpconf}
\usepackage{graphicx}
\begin{document}
\title{Hadronic matter phases and their application to rapidly rotating neutron stars}

\author{Tomoki Endo}

\address{Division of Physics, Department of General Education, National Institute of Technology, Kagawa College, 355 Chokushi-cho, Takamatsu, Kagawa 761-8058, Japan}

\ead{endo@t.kagawa-nct.ac.jp}

\begin{abstract}
Neutron stars are commonly considered as astronomical objects having high-density interiors and an inner core region in which various hadronic matter phases are expected. Several studies show that the inner structures affect macroscopic phenomena of the star. However, we know that the inner structures of the star strongly depend on the equation of state (EOS). The EOS of high-density matter is still not clear and several recent observations indicate restrictions to EOSs. Theoretical studies should elucidate EOSs at high density and/or high temperature. For instance, many theoretical studies have attempted to account for the rotation effect of rapidly rotating neutron stars (i.e., pulsars). Accordingly, we also apply our EOSs to rapidly rotating stars. Furthermore, neutron stars generate a strong magnetic field. Several recent studies indicate that this magnetic field exerts restrictions on the EOS. In this paper, we focus on the investigation of the inner structures and the application of our EOSs to rotating stars. We find that one of our EOSs is consistent with observations, and another is inconsistent. We also find an important relation between the radius and rotation.
\end{abstract}

\section{Introduction}
Neutron stars are often described as ``astrophysical laboratories'' because they display diverse physical phenomena. The stars have extremely high density, strong magnetic fields, strong gravity and rapid rotation. Stars that rapidly rotate at several hundred hertz are called ``millisecond pulsars''.  It is believed that exotic physical phenomena appear in and/or around such stars. The mass--radius relation is an important feature of neutron stars and has been described in theoretical studies by the Tolman--Oppenheimer--Volkoff (TOV) equation\cite{shapiro}:
\begin{eqnarray}
\frac{dp(r)}{dr} &=& -\frac{\left[p(r)+\epsilon(r)\right] \left[M(r)+4\pi
			  r^3 p(r)\right]}{r \left[r-2M(r)\right]} \\
M(r) &\equiv& 4 \pi \int_0^r \epsilon (r) r^2 dr.
\label{tov}
\end{eqnarray}
 However, several recent studies have suggested limitations of the theoretical studies. In particular, the TOV equation assumes that the star is spherical.  Neutron stars are observed as pulsars, which are known to have fast rotation. Such stars should be elliptical rather than spherical because of the rotation effect \cite{belv}. 

 To investigate the properties of a neutron star, we need the equation of state (EOS) for high-density matter. In particular, theoretical studies are attempting to elucidate the EOS for quark matter because the central density of the star is sufficiently high that nuclear matter becomes quark matter. Currently accepted theories and many experimental results suggest that hadronic matter changes to quark matter in high-density and/or high-temperature regimes by way of the deconfinement phase transition. The properties of quark matter have been 
 actively studied theoretically in terms of the quark--gluon
 plasma, color superconductivity \cite{alf1,alf3}, and magnetism \cite{tat1,tat3,dex1},
 and experimentally in terms of relativistic
 heavy-ion collisions \cite{rhic}, the early universe and compact stars
 \cite{mad3,chen,bejg,mini}. 
 Such studies are continuing to provide exciting results \cite{risch}.
 Presently, we consider that compact stars consist of not only nuclear matter but also other matter such as hyperons and quarks. We call such stars {\it hybrid stars}.

Because many theoretical calculations have suggested that the deconfinement 
phase transition is of the first order
at low temperature and high density \cite{pisa,latt},  
 we assume that it is a first-order phase transition here.
 The Gibbs conditions \cite{gle1}
 then give rise to various structured mixed phases.
 The structured mixed phases proposed by Heiselberg et al.\ \cite{pet} and Glenndening
 and Pei \cite{gle2} suggest a crystalline structure for the mixed phase
 in the cores of hybrid stars. One phase is embedded in the another phase with geometrical structures. Such structures are
 called ``droplets'', ``rods'', ``slabs'', ``tubes'', and ``bubbles''. 
 We present the EOS for the mixed phase taking
 into account the charge screening effect \cite{end2} without relying on 
 any approximations. We investigate the inner structures of these stars \cite{end3}. In this paper, we apply our EOS to a stationary rotating star.

\section{Formalism}
Our EOS was presented in detail in Ref.\ \cite{end2,end3} and here we only briefly review the EOS.
Our models are described as follows. The quark matter has three flavors: {\it u}, {\it d}, and {\it s} quarks. Additionally,
the electron is in the quark phase. We incorporate the MIT bag model and assume a sharp boundary at the
hadron--quark interface. 
{\it u} and {\it d}
quarks are treated as massless and {\it s} as being massive
($m_s=150$MeV), and the quarks interact via the one-gluon-exchange interaction.
Hadron matter comprises the proton, neutron and electron, which constitute simple nuclear matter. We use the effective potential to reproduce the saturation properties of the nuclear matter.

At the phase transition, we maintain strict thermodynamic conditions; i.e., the Gibbs conditions':
\begin{equation}
 \mu_{\mathrm{B}}^{\mathrm{Q}} = \mu_{\mathrm{B}}^{\mathrm{H}} (\equiv \mu_{\mathrm{B}} ),
 \hspace{5pt} \mu_{\mathrm{charge}}^{\mathrm{Q}} =
 \mu_{\mathrm{charge}}^{\mathrm{H}},  \hspace{5pt} P^{\mathrm{Q}}=P^{\mathrm{H}}, \hspace{5pt} T^{\mathrm{Q}} = T^{\mathrm{H}},
\label{gc}
\end{equation}
where the superscript H(Q) denotes the hadron (quark) phase, and
$\mu_{\mathrm{B}}^\mathrm{H(Q)}$ and
$\mu_{\mathrm{charge}}^\mathrm{H(Q)}$  are the baryon number
and charge chemical potentials, respectively. 
Note that there are two independent chemical potentials in this phase transition.
Such a case should be much different from the liquid--vapor 
phase transition that is described by a single chemical potential. 

Furthermore, we need to use the thermodynamic potential in the calculation of the phase transition.
The total thermodynamic potential ($\Omega_\mathrm{total}$) 
consists of hadron, quark and electron contributions and the
surface contribution:
$\Omega_\mathrm{total} = \Omega_\mathrm{hadron} +\Omega_\mathrm{quark}
 +\Omega_\mathrm{surface},$ 
where $\Omega_\mathrm{hadron(quark)}$ denotes the contribution of the hadron (quark)
phase. We here introduce the surface contribution $\Omega_\mathrm{surface}$,
parameterized by the surface tension value $\sigma$, $\Omega_\mathrm{surface} = \sigma S$, with 
$S$ being the area of the interface.
Note that $\Omega_\mathrm{surface}$ may be closely related with the deconfinement problem and unfortunately 
there is no known method that adequately treats the problem. 
Therefore, many authors have treated the strength as a parameter and investigated 
how its value affects the results; the values used range 10--100 $\mathrm{MeV/fm}^2$ \cite{pet,gle2,alf2}.
 We take the value of 40 $\mathrm{MeV/fm}^2$, which is a moderate value, in our calculation. 
To investigate the charge screening effect, we also make calculations 
without the screening effect \cite{end2, maru, maru2}. 
We then apply the EOSs derived in our paper \cite{end2} to
the TOV equation \cite{end3,end4}. 

We finally apply our EOS to a stationary rotating star.
However, it is difficult to take into account the rotation effect of relativistic stars. We therefore assume:

1) stationary rigid rotation (i.e., uniform rotation),

2) axial symmetry with respect to the spin axis, and

3) the matter being a perfect fluid.

Stationary rotation in general relativity has been reviewed in \cite{ster} and \cite{kurk}; we follow their calculation. We then apply our EOS to a stationary rotating star.

\section{Numerical results}

\begin{figure}[htb]
\begin{center}
\includegraphics[width=75mm]{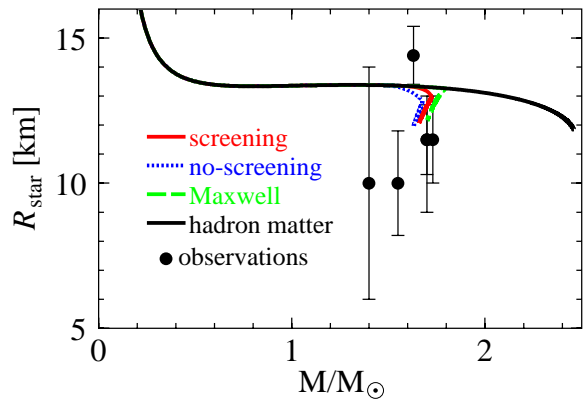}
\caption{(Color online) Mass--radius relation of stars obtained with our models plotted against the observational data listed in \cite{kurk}. The difference between screening and no-screening mixed phases is clearly small.}
\label{M-R}
\end{center}
\end{figure}

We first review the result of solving the TOV equation using our EOSs. Figure\ \ref{M-R} shows the mass--radius relations of stars with and without screening, using the Maxwell construction, and in the case of pure hadronic matter. As done in our former paper \cite{end3}, we note the error bar of the radius.  

\begin{figure}[htb]
\begin{center}
\includegraphics[width=75mm]{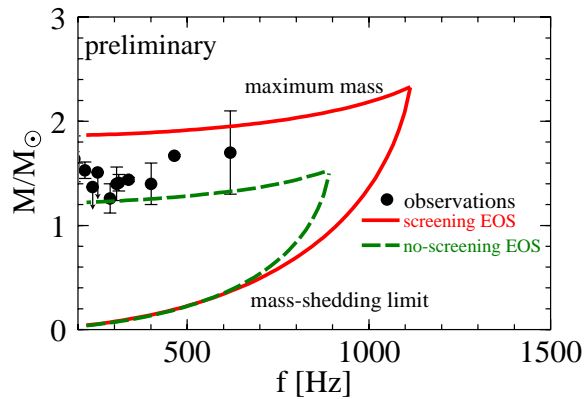}
\caption{ (Color online) Mass--frequency relation obtained with our models plotted against the observational data listed in \cite{kurk}. The solid and dashed curves represent the results obtained with and without screening, respectively.}
\label{M-f}
\end{center}
\end{figure}

Figure\ \ref{M-f} shows the result for a rotating star obtained using our EOSs with and without screening. The upper curve shows the maximum mass of the star and the lower curve shows the mass-shedding curve, which corresponds to the Kepler frequency. The Kepler frequency indicates that the centrifugal force is equal in magnitude to gravity. Therefore, the area on the right-hand side of the lower curve is physically invalid. Black dots are observations. Then, if the upper curve is lower than the observations, the EOS should be ruled out. Our EOS in the screening case is thus consistent with these observations. However, our EOS without screening is not consistent and therefore inappropriate. This difference could be derived by the softness of the EOS \cite{kurk}, although further investigations are needed.

\begin{figure}[htb]
\begin{center}
\includegraphics[width=75mm]{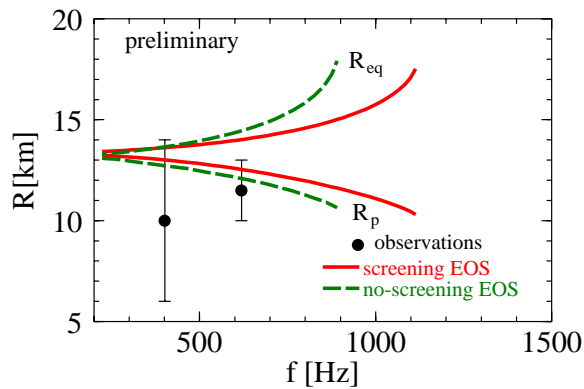}
\caption{ (Color online) Radius--frequency relation of our model plotted against observational data (SAXJ1808.4-3658 and 4U1608-52). The solid and dashed curves represent the results obtained with and without screening, respectively.}
\label{R-f}
\end{center}
\end{figure}

Figure\ \ref{R-f} shows an important relation between the radius and rotation. The radius of a star is considered a single value because we ordinarily consider a star that is approximately spherical. However, if the star is rapidly rotating, it should be elliptical. We thus have to recognize the different radii.
According to assumption 2), we have to consider two different radii, $R_{\mathrm eq}$ and $R_{\mathrm p}$, which are the equatorial radius and polar radius, respectively. Figure\ \ref{R-f} shows $R_{\mathrm eq}$ and $R_{\mathrm p}$ with respect to rotation. If the rotation rate is 400 Hz or higher, the two radii are different. We thus have to note the effects of rotation on rapidly rotating stars.

\section{Summary and Concluding Remarks} \indent

In this paper, we presented the difference between EOSs with and without the charge screening effect while taking into account rotation effects.
We found that the EOS in the screening case could reproduce the maximum mass of the observed data while that without screening could not with respect to the spinning. Furthermore, the shape of the star should be affected by the rotation, and we then have to consider two different radii. Therefore, when studying rapidly rotating neutron stars, we have to pay attention to the ``radius''.
Moreover, we need to further improve our models. In this study, we used a simple model for quark matter and nuclear matter.
To obtain a more realistic picture of the hadron--quark phase transition, we need to
take into account color superconductivity \cite{alf1,alf2} and relativistic mean field theory \cite{shen}. We will then be able to obtain results that are more realistic. 
 Strong magnetic fields are a well-known and attractive physical phenomenon of neutron stars \cite{orsa1,orsa2,webe}. However, the origin of these strong magnetic fields is still not clear. Theoretical studies have attempted to explain the strong magnetic fields using various methods \cite{rabh,sinh,grac,grun,isay}. One possible explanation if the spin-polarization of quark matter \cite{tat3,tat4,tsue1,tsue2}. However, whether quark matter exists or not strongly depends on the EOS of the matter.
 We did not take into account magnetic fields in this study. If a magnetic field is included in our calculation, it will be possible to investigate the relation between the magnetic field and the rotation effects \cite{chir}.

\section*{Acknowledgments} \indent

This work was supported in part by the Principal Grant of the National Institute of Technology, Kagawa College.


\section*{References}

\begin{thebibliography}{9}
\bibitem{shapiro} Shapiro S L and Teukolsky S A, {\it Black holes,
	white dwarfs, and neutron stars, the physics of compact objects} Wiley-interscience

\bibitem{belv} Belvedere R, Boshkayev K, Rueda J A, and Ruffini R 2014 {\it Nucl. Phys.} A {\bf 921} 33

\bibitem{alf1} Alford M, Schmitt A,
	Rajagopal K and Sch\"{a}fer T 2008 {\it Rev. Mord. Phys.} {\bf 80} 1455 and references therein

\bibitem{alf3} Alford M and Reddy S 2003 {\it Phys. Rev.} D {\bf 67} 074024

\bibitem{tat1} Tatsumi T, Maruyama T and Nakano E 2004 {\it Prog. Theor. Phys. Suppl.} {\bf 153} 190

\bibitem{tat3} Tatsumi T 2000 {\it Phys. Lett.} B {\bf 489} 280

\bibitem{dex1} Dexheimer V, Negreiros R and Schramm S, 2012 {\it  Eur.Phys.J.} {\bf A48} 189

\bibitem{rhic} Adcox K, et al.\ (PHENIX collaboration) 2002
	{\it Phys. Rev. Lett.} {\bf 88} 022301; Adler C, et al.\ (STAR
	collaboration) 2003 {\it Phys. Rev. Lett.} {\bf 90} 082302

\bibitem{mad3} Madsen J 1999 {\it Lect. Notes Phys.} {\bf 516} 162

\bibitem{chen} Cheng K S, Dai Z G and Lu T 1998 {\it Int. Mod. Phys.} D {\bf 7}
	139
\bibitem{bejg}  Bejger M, Haensel P and Zdunik J L 2005 {\it Mon. Not. Roy. Astron. Soc.} {\bf 359} 699

\bibitem{mini} Minuitti G, Pons J A, Berti E, Gualtieri L and
	Ferrari V 2003 {\it Mon. Not. Roy. Astron. Soc.} {\bf 338} 389

\bibitem{risch} Rischke D H 2004 {\it Prog. Part. Nucl. Phys.} {\bf 52} 197

\bibitem{pisa} Pisalski R D and Wilczek F 1984
	{\it Phys. Rev. Lett.} {\bf 29} 338

\bibitem{latt} Gavai R V, Potvin J and Sanielevici S 1987
	{\it Phys. Rev. Lett.} {\bf 58} 2519.

\bibitem{gle1} Glendenning N K 1992 {\it Phys. Rev.} D {\bf 46} 1274; 
2001 {\it Phys. Rep.} {\bf 342} 393 

\bibitem{pet} Heiselberg H, Pethick C J and Staubo E F 1993 {\it Phys. Rev. Lett.} {\bf 70} 1355

\bibitem{gle2} Glendenning N K and Pei S 1995 {\it  Phys. Rev.} C {\bf 52} 2250

\bibitem{end2}  Endo T, Maruyama T, Chiba S and Tatsumi T 
	2006 {\it Prog. Theor. Phys.} {\bf 115} 337

\bibitem{end3}  Endo T 2011 {\it Phys. Rev.} C {\bf 83} 068801


\bibitem{alf2} Alford M, Rajagopal K, Reddy S, and Wilczek F 2001 {\it Phys. Rev.} D {\bf 64} 074017

\bibitem{maru}  Maruyama T, Tatsumi T, Voskresensky D N,
	Tanigawa T, Endo T and Chiba S 2006 {\it Phys. Rev.} C {\bf 73} 035802

\bibitem{maru2}  Maruyama T, Tatsumi T, Endo T and Chiba S 2006 {\it Recent Res. Devel. Phys.} {\bf 7} 1; nucl-th/0605075

\bibitem{end4} Endo T arXiv:1310.0913[astro-ph.HE]

\bibitem{ster} Stergioulas N, Friedman J L 1995 {\it  Astrophys. J} {\bf 444} 306 

\bibitem{kurk} Kurkela A, Romatschke P, Vuorinen A and Wu B arXiv:1006.4062[astro-ph.HE]

\bibitem{shen} Shen H, Toki H, Oyamatsu K and Sumiyoshi K 1998 {\it Nucl. Phys.} A {\bf 637} 435 

\bibitem{orsa1} Orsaria M, Rodrigues H, Weber F and Contrera G A 2013 {\it Phys. Rev.} D {\bf 87} 023001

\bibitem{orsa2} Orsaria M, Rodrigues H, Weber F and Contrera G A 2014 {\it Phys. Rev.} C {\bf 89} 015806

\bibitem{webe} Weber F, Orsaria M, Negreiros R arXiv:1307.1103[astro-ph.SR]

\bibitem{rabh} Rabhi A and Provid$\hat{\rm{e}}$ncia C 2011 {\it Phys. Rev.} C {\bf 83} 055801 

\bibitem{sinh} Sinha M, Huang X and Sedrakian A 2013 {\it Phys. Rev.} D {\bf 88} 025008

\bibitem{grac} Gracia A F and Pinto M B 2013 {\it Phys. Rev.} C {\bf 88} 025207

\bibitem{grun} Grunfeld A G, Menezes D P Pinto M B and Scoccola N N 2014 {\it Phys. Rev.} D {\bf 90} 044024

\bibitem{isay} Isayev A A 2015 {\it Phys. Rev.} C {\bf 91} 015208

\bibitem{tat4} Tatsumi T arXiv:1107.0807[hep-ph]

\bibitem{tsue1} Tsue Y, Provid$\hat{\rm{e}}$ncia J, Provid$\hat{\rm{e}}$nsia C, Yamamura M and Bohr H 2015 {\it Prog. Theor. Exp. Phys.} {\bf 1} 013D02

\bibitem{tsue2} Tsue Y, Provid$\hat{\rm{e}}$ncia J, Provid$\hat{\rm{e}}$nsia C, Yamamura M and Bohr H 2015 {\it Prog. Theor. Exp. Phys.} {\bf 10} 103D01

\bibitem{chir} Chirenti C and Sk$\acute{\rm a}$kala J 2013 {\it Phys. Rev.} D {\bf 88} 104018

\end{thebibliography}
\end{document}